\documentclass[aps,prb,twocolumn,showpacs,floatfix]{revtex4-1}
\usepackage{graphicx}
\usepackage{dcolumn}
\usepackage{bm}
\usepackage{natbib}
\usepackage{caption}
\usepackage{subcaption}
\usepackage{float}
\usepackage{amsmath,amssymb}

\newcommand{\comment}[1]{}

\newcommand{\subscript}[1]{\ensuremath{_{\textrm{#1}}}}

\begin{document}
\title{First Principles Calculation of Elastic Moduli of Early-Late Transition Metal Alloys}
\author{William Paul Huhn}
\affiliation{Department of Physics, Carnegie Mellon University, Pittsburgh, PA 15213}
\author{Michael Widom}
\affiliation{Department of Physics, Carnegie Mellon University, Pittsburgh, PA 15213}
\author{Andrew M. Cheung}
\affiliation{Department of Materials Science and Engineering, University of Virginia, Charlottesville, VA 22904}
\author{S. Joseph Poon}
\affiliation{Department of Physics, University of Virginia, Charlottesville, VA 22904}
\author{Gary J. Shiflet}
\affiliation{Department of Materials Science and Engineering, University of Virginia, Charlottesville, VA 22904}
\author {John Lewandowski}
\affiliation{Department of Materials Science and Engineering, Case Western Reserve University, Cleveland, OH 44106}
\begin{abstract}
Motivated by interest in the elastic properties of high strength amorphous metals, we examine the elastic properties of select crystalline phases.  Using first principles methods, we calculate elastic moduli in various chemical systems containing transition metals, specifically early (Ta,W) and late (Co,Ni).  Theoretically predicted alloy elastic properties are verified for Ni-Ta by comparison with experimental measurements using resonant ultrasound spectroscopy. Comparison of our computed elastic moduli with effective medium theories shows that alloying leads to enhancement of bulk moduli relative to averages of the pure elements, and considerable deviation of predicted and computed shear moduli.  Specifically, we find an enhancement of bulk modulus relative to effective medium theory and propose a candidate system for high strength, ductile amorphous alloys.  Trends in the elastic properties of chemical systems are analyzed using force constants, electronic densities of state and Crystal Overlap Hamilton Populations.  We interpret our findings in terms of the electronic structure of the alloys.
\end{abstract} 
\maketitle

\section{Introduction}
Elastic moduli are important for understanding various properties
of amorphous metals.  Bulk moduli, shear moduli, and their ratio
correlate to glass transition temperature~\cite{Egami97,Egami07},
glass forming ability~\cite{WHWang06},
brittleness~\cite{Poon08,HSChen75,JJLewandowski05}, Gr\"{u}neisen
parameters~\cite{JQWang09}, maximum resolved shear stress at
yielding~\cite{WLJohnson05}, chemical bonding type~\cite{Wang12}, and
possibly fragility~\cite{Wang12}.  Knowledge of the bulk and shear
moduli is thus important for materials design.  However, amorphous
materials commonly contain at least 4 chemical species, making exhaustive
experimental evaluation of candidate materials impossible.  Empirical
methods for predicting stoichiometries with desired
properties are therefore necessary.

First principles computational methods prove fruitful, owing to their
chemical specificity and absence of adjustable parameters, as well as
the insight they yield into electronic structure.  However, amorphous
metals pose computational difficulties, as they lack both spatial
periodicity and a unique structure.  While the first problem can be
practically overcome by imposing suitably large periodic boundary
conditions, this requires hundreds of atoms per computational cell,
straining computational resources, and requiring averaging over
multiple samples to remove sample dependence.  The second problem can
be partially overcome by running molecular dynamics on a liquid sample
then rapidly quenching the sample. However the requirement for
equilibration further increases the computational time necessary.
Hence we adopt a different strategy.

\begin{figure*}
\begin{subfigure}[b]{0.18\textwidth}
\includegraphics[trim = 0mm -1mm 0mm 0mm, clip, height=1in]{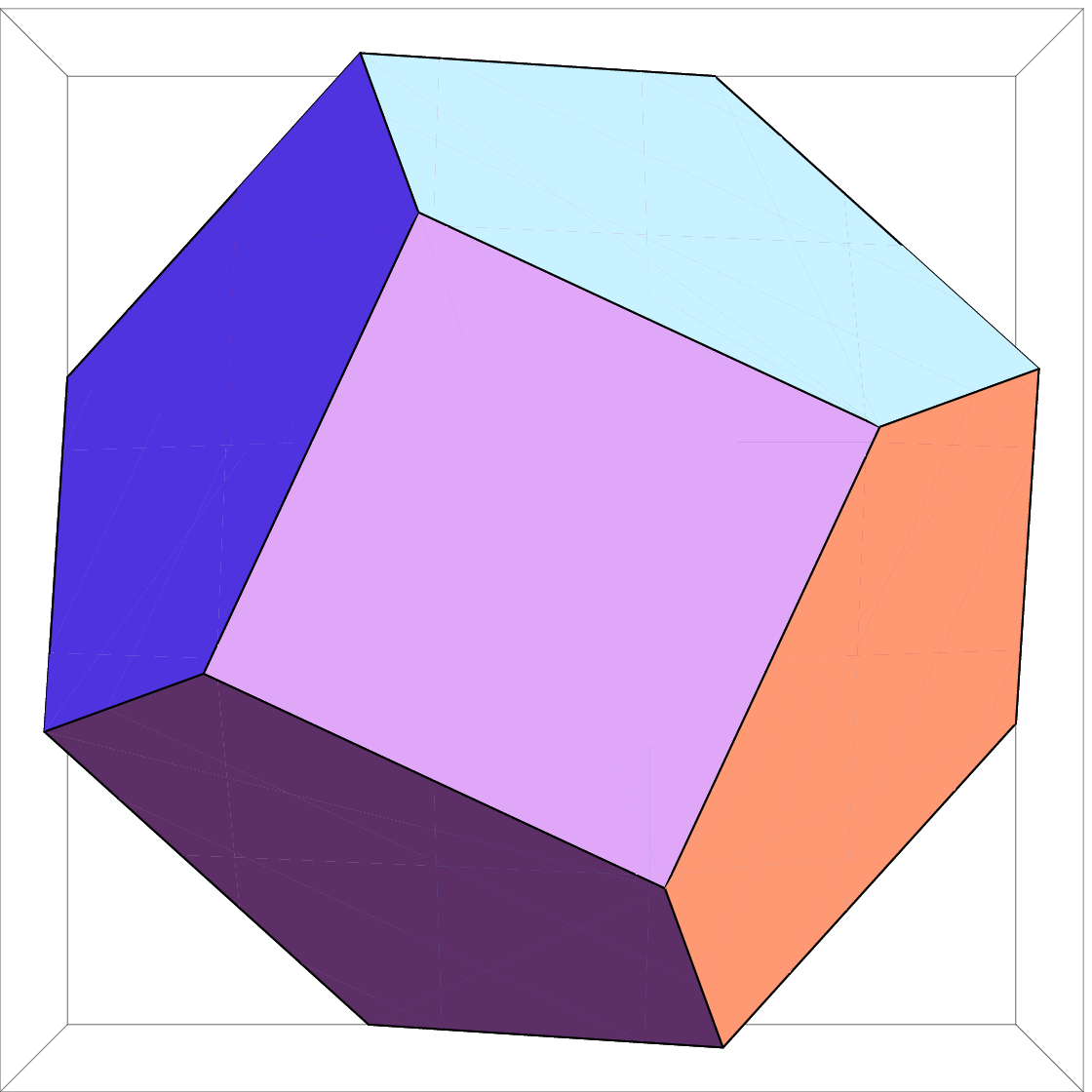}
\caption{CN=10, (0,2,8)}\end{subfigure}
\begin{subfigure}[b]{0.18\textwidth}
\includegraphics[trim = 0mm -1mm 0mm 0mm, clip, height=1in]{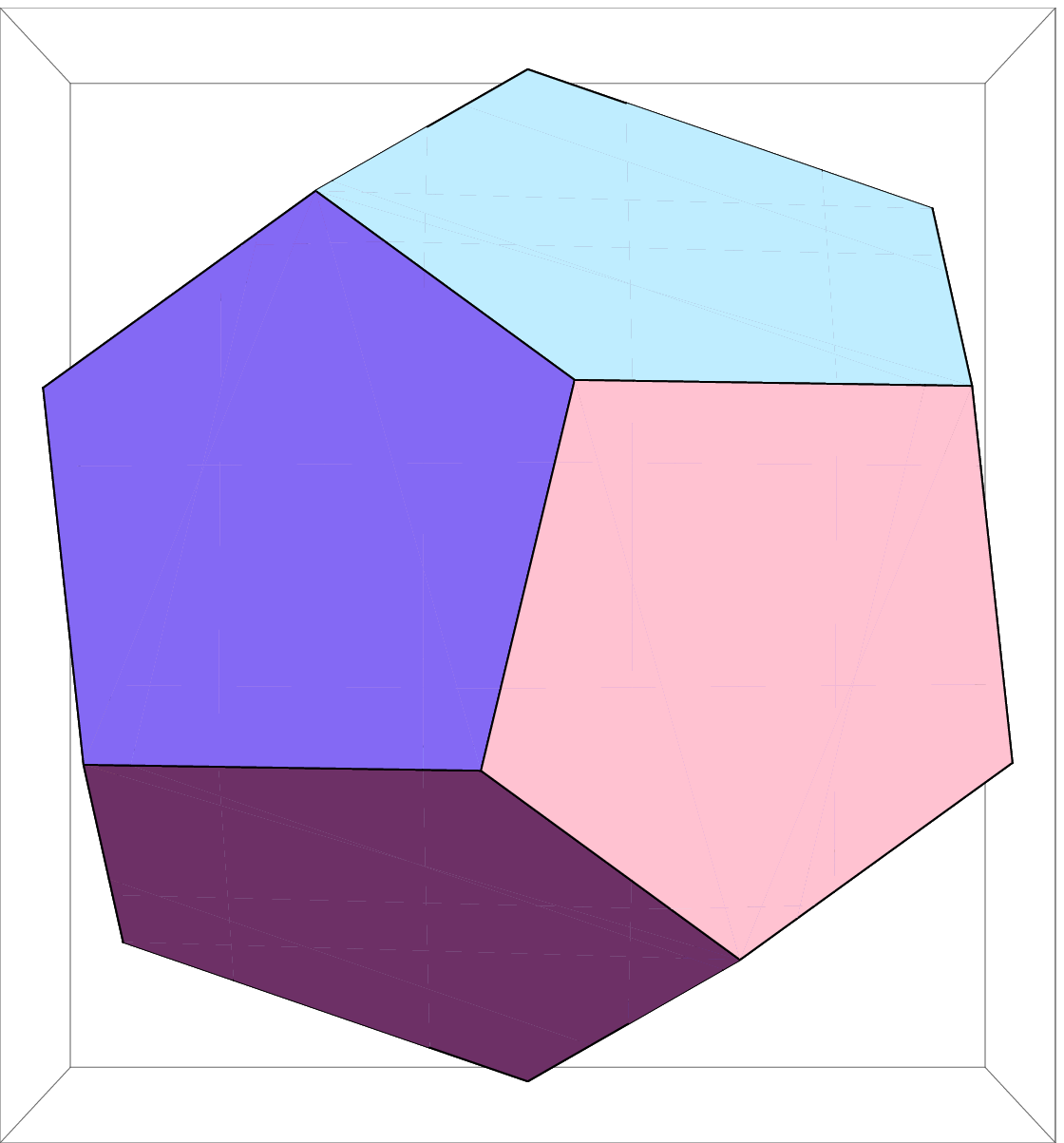}
\caption{CN=12, (0,0,12)}\end{subfigure}
\begin{subfigure}[b]{0.18\textwidth}
\includegraphics[trim = 0mm -1mm 0mm 0mm, clip, height=1in]{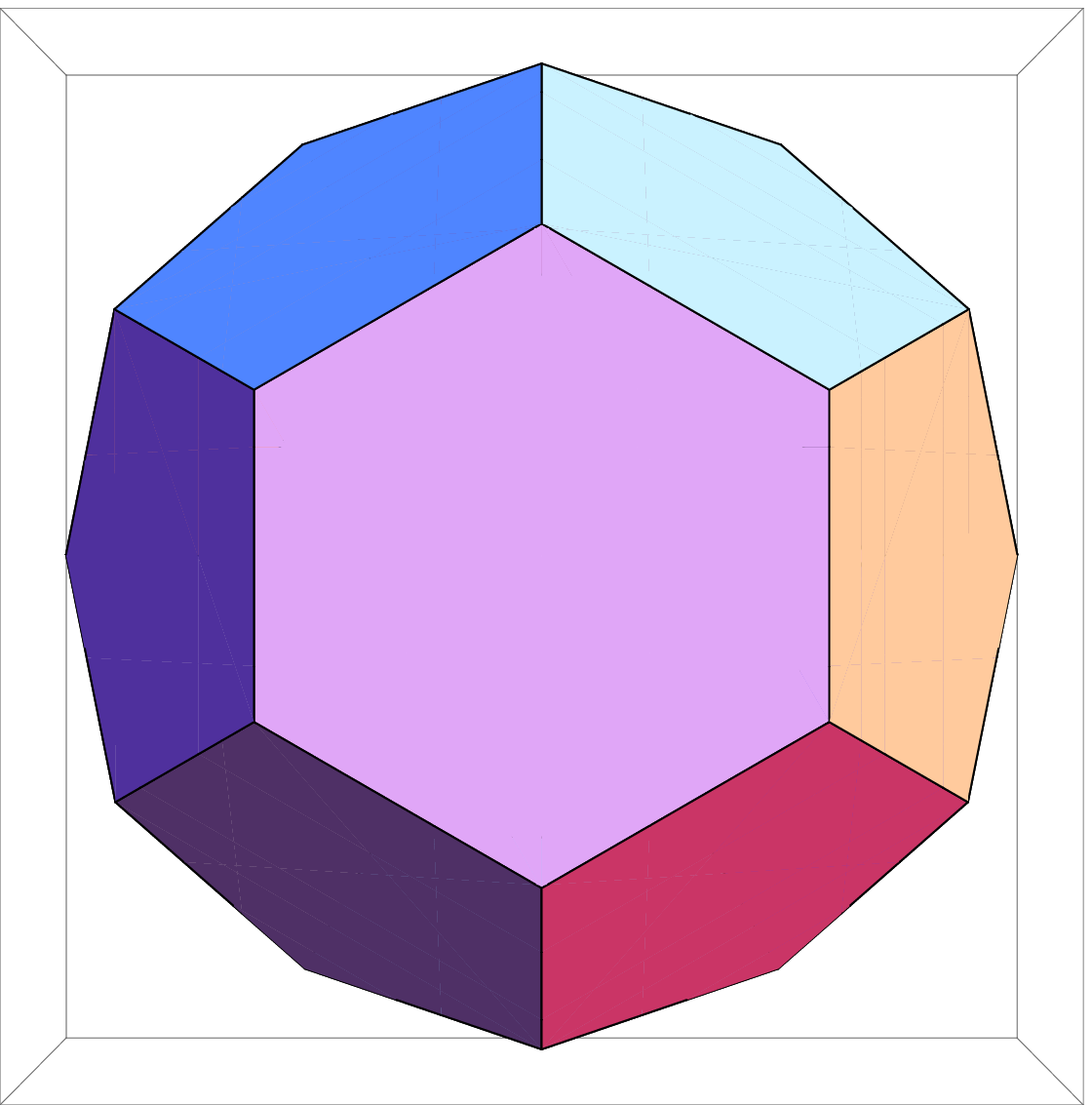}
\caption{CN=14, (0,0,12,2)}\end{subfigure}
\begin{subfigure}[b]{0.18\textwidth}
\includegraphics[trim = 0mm -1mm 0mm 0mm, clip, height=1in]{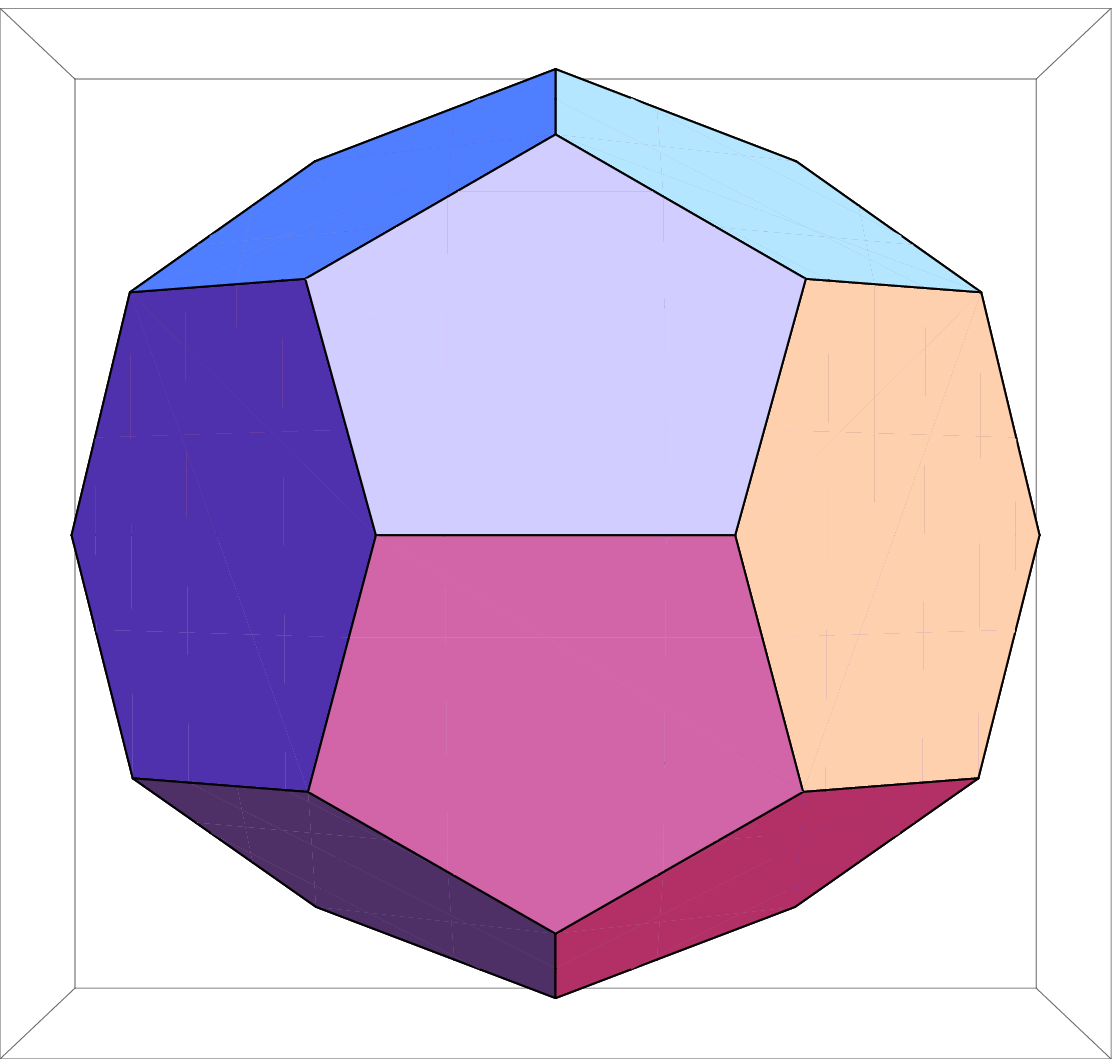}
\caption{CN=15, (0,0,12,3)}\end{subfigure}
\begin{subfigure}[b]{0.18\textwidth}
\includegraphics[trim = 0mm -1mm 0mm 0mm, clip, height=1in]{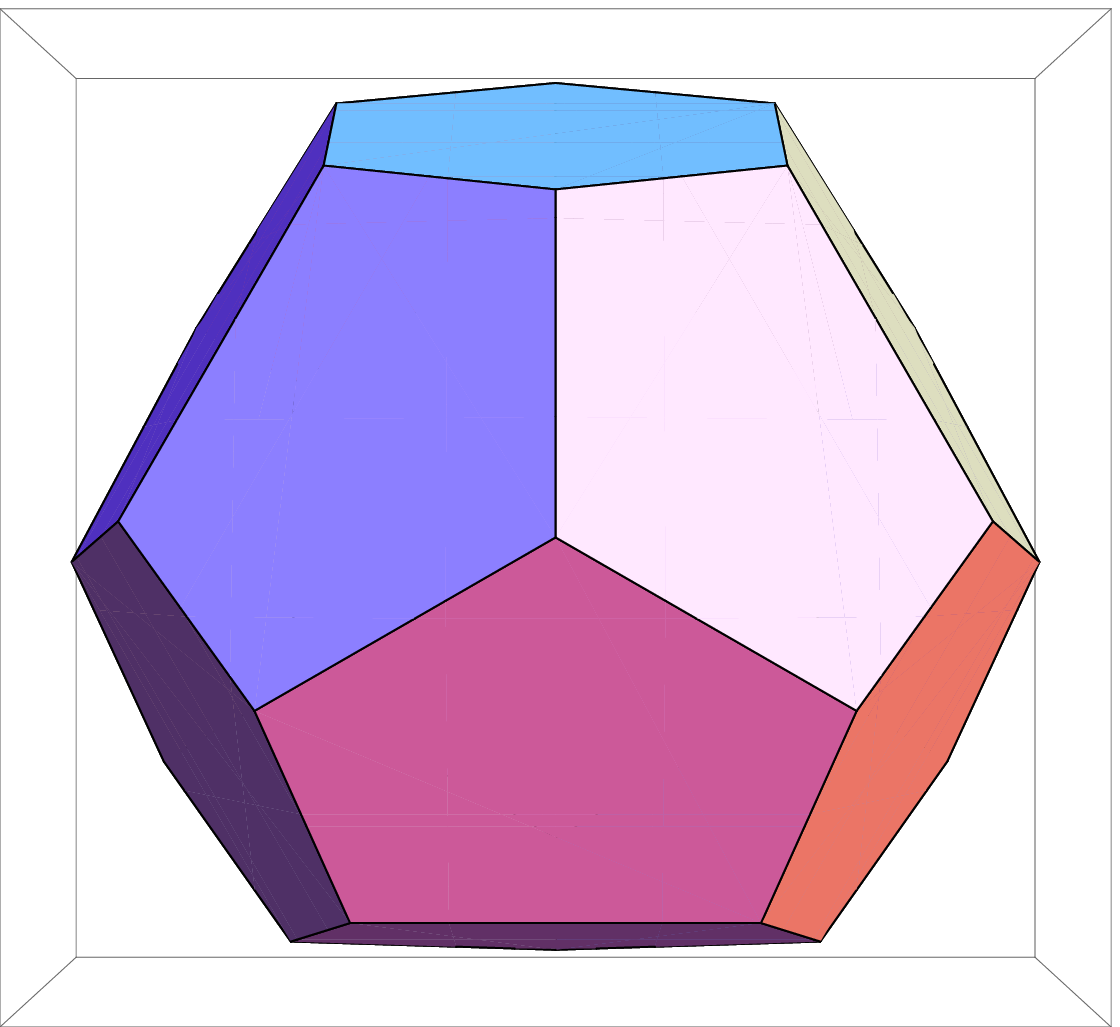}
\caption{CN=16, (0,0,12,4)}\end{subfigure}
\caption{\label{fig:FK}Frank-Kasper Voronoi polyhedra, listing coordination numbers and polyhedron codes.}
\end{figure*}

Frank Kasper phases~\cite{Frank58,Frank59,Sinha72} are complex but
otherwise ordinary crystalline phases.  Due to their topological close
packing they exhibit local icosahedral ordering similar to that found
in many amorphous metals.  Fig.~\ref{fig:FK} shows the standard
Voronoi polyhedra of Frank-Kasper structures.  We expect that the
similar local chemical environments of the Frank Kasper phases can be
used to mimic amorphous metals, yielding ``amorphous approximants'',
similar in concept to ``quasicrystal approximants''.  These
crystalline phases will be used to understand trends in the elastic
properties of amorphous metals. It is observed that in amorphous
metals, shear moduli are typically 20\%-30\% lower, and bulk moduli 
5\%-10\% lower, compared to crystalline phases of similar
composition~\cite{HSChen75}.  Many crystalline phases have small unit
cells compared to system sizes required to reproduce amorphous
structures.

Amorphous metals can exhibit high strength~\cite{Telford200436}, but
often at the cost of a lack of ductility owing to the absence of
dislocations.  Designing amorphous alloys for high ductility is
impeded both by the challenge of formulating accurate amorphous
structural models, and by the lack of a valid predictive theory of
ductility even for the case of crystalline compounds, though there are
empirical rules based on the Poisson ratio, or equivalently the
shear/bulk modulus ratio.  We hypothesize that relatively ductile
crystalline compounds will tend to create relatively ductile amorphous
compounds.  Further, metallic glass composites, consisting of
crystalline grains embedded in an amorphous matrix, have been shown to
increase the toughness, impact resistance and plastic strain to
failure~\cite{PhysRevLett.84.2901}, further motivating the
investigation into elastic properties of crystalline phases in
metallic glass-forming alloy systems.

\section{Background and Methods}
\subsection{Elasticity}
The fundamental equation of linearized elasticity is
$\sigma_i=C_{ij}\epsilon_j$, where $\sigma$ is the stress tensor,
$\epsilon$ is the strain tensor, and $C$ the stiffness tensor relating
the two.  Here we employ Voigt notation, converting the tensors
$\sigma$ and $\epsilon$ into vectors with 6 components
$\{xx,yy,zz,xy,xz,yz\}$.  For crystals $C$ depends on the symmetry of
the material in question~\cite{Landau86} and contains at least three independent
parameters.  For the special case of isotropic materials $C$ has only
two independent parameters, the bulk modulus $K$ and the shear modulus
$G$.  The non-vanishing elements are
$C_{11}=C_{22}=C_{33}=K+4G/3$, $C_{12}=C_{13}=C_{23}=K-2G/3$, and
$C_{44}=C_{55}=C_{66}=G$.  An additional elastic quantity of interest
is the Poisson ratio $\nu$, defined to be the negative of the ratio of
axial strain to transverse strain.  For crystals $\nu$ depends on the
direction of applied stress, but in the isotropic case it reduces to
\begin{equation}
\nu=\frac{3-2G/K}{6+2G/K}.
\label{eq:poisson}
\end{equation}
Notice that $\nu$ is a function of the shear to bulk modulus ratio 
$G/K$, also known as the Pugh ratio~\cite{Pugh54}.

Ductility and Poisson's ratio are positively correlated in both
polycrystalline~\cite{Pugh54} and amorphous~\cite{Poon08} solids, with
$\nu\geq0.32$ (equivalently $G/K\leq0.41$) proposed as a empirical
criterion for good ductility ~\cite{Gu06}.  Seeking simultaneously
high strength and ductility, it is natural to choose chemical species
which in elemental form have large $K$.  The species chosen must also
be known good glass formers.  Early-late transition metal alloys fit
both criteria, as the transition metals from Group IV to XI all have 
$K\geq100$ GPa, and they are one of the most frequently examined
classes of amorphous metals, .  Size mismatch criteria favor using
transition metals from different rows and columns of the periodic
table, e.g. Ta and W for early transition metals (ETMs), and Co and Ni
as late transition metals (LTMs).  Co and W are of particular interest
for materials design as Co-based glasses exhibit ultrahigh fracture
strength, W-based glasses have the highest known glass transition 
temperature for bulk metallic glasses, and both have high Young's 
modulus~\cite{AInoue03, Wang12, MOhtsuki04}.

Ordinary crystalline materials contain randomly orientated microscopic
grains and appear macroscopically isotropic.  To compute the elastic
properties for these ``polycrystals'', orientational averaging is
required.  Each grain is described microscopically by the same
stiffness matrix, which contains between 3 (cubic) and 21 (triclinic)
independent parameters~\cite{Nye85}, but macroscopically the crystal
is isotropic with only two independent parameters $K$ and $G$.  In the
Voigt average, the stiffness matrix $C$ is averaged over
orientations~\cite{Zeller73}, which is exact if stress is uniform
throughout space.  In the Reuss average, the compliance matrix
$S=C^{-1}$ is averaged over orientations, which is exact if strain is
uniform throughout space.  The Voigt average systematically
overestimates the isotropic moduli, while the Reuss average
systematically underestimates the moduli.  Empirically, the arithmetic
mean of the two, known as the Hill average~\cite{Hill52}, gives
improved agreement with experiment, and this is what we will report.

The polycrystalline average assumes the stiffness matrices of various
grains are identical to the perfect crystal, differing only by
relative orientation.  To obtain macroscopic elastic moduli in
materials where chemical environment spatially varies, in particular
in materials containing chemically distinct grains, it is necessary to
include effects arising from fluctuations in the local stiffness
matrices.  A general class of approximation schemes known as
``effective medium theories'' exists where each grain with its local
stiffness matrix interacts with an effective medium characterized by a
background stiffness matrix incorporating the interactions of all
other grains.  A popular effective medium theory is the coherent
potential approximation (CPA)~\cite{Nan93}, a self-consistent
formalism in which the background stiffness matrix is taken as the
macroscopic average itself.  The self-consistent interaction of grains
yields a pair of coupled non-linear equations,
\begin{equation}
\sum_{i} \phi_{i}\frac{K_{i}-K}{3K_{i}+4G} = 0
\label{eq:CPA_K}
\end{equation}
and 
\begin{equation}
\sum_{i} \phi_{i}\frac{G_{i}-G}{5G(3K+4G)+6(K+2G)(G_{i}-G)} = 0,
\label{eq:CPA_G}
\end{equation}
which can be solved numerically for the effective $K$ and $G$, where
$\phi_{i}$ denotes the volume fraction of grain type $i$ in the
material. $K_{i}$ and $G_{i}$ denote the bulk and shear modulus of
grain type $i$, respectively.
 
Although intended for mixtures of crystalline grains, we will apply
CPA in the limit where each grain shrinks to a single atom, to
estimate the elastic moduli of compounds using the self-consistent
average of properties of the constituent pure elements.  Our usage of
CPA may be viewed as a higher-order correction to the well-known but
highly empirical ``rule of mixtures'' paradigm common in materials
design, which has already been applied to bulk metallic
glasses~\cite{YZhang07}.  CPA takes into account the pure elemental
properties but lacks information about interspecies bonding and
specific alloy crystal structure.  Thus we take CPA as a convenient,
physically motivated interpolation to establish a baseline for
comparison with the computed alloy moduli revealing the specific
contributions of structure and bonding, though we note that metallic
glasses exist where trace changes in composition yield relatively
large changes in K and G due to alteration of chemical bonding
type~\cite{BZhang06}.

\subsection{First principles methods}

Our first principles calculations use the Vienna Ab-Initio Simulation
Package (VASP)~\cite{KresseHafner93,KresseFurthmuller96}, a plane wave
ab-initio package implementing PAW pseudopotentials~\cite{Blochl94} in
the PW91~\cite{PerdewWang92} generalized gradient approximation to
density functional theory (DFT).  VASP calculates total energies,
forces, elastic moduli, and electronic structure.  All structures are
relaxed until the maximum ionic force is below 0.01 eV/\AA, and the
k-point density is subsequently increased until total energy per atom
has converged to within 0.1 meV/atom.  Default plane wave energy
cutoffs are used for total energy calculations.  In structures
containing Co and Ni, spin polarization has been included.  However,
we do not include spin-orbit coupling despite the presence of 5d
elements Ta and W.  Total energies are converted to enthalpies of
formation by subtracting from the tie-line joining the total energies
of pure elements in their stable crystalline forms~\cite{Mihal04}.

We perform elastic calculations using a finite difference method
internal to VASP.  To ensure proper convergence of elastic moduli, we
increase the k-point density until the all polycrystalline averages
converge to within 2\%, then increase the energy cutoff until the
polycrystalline averages converge.  Convergence in energy cutoff
occurs at 360 eV, with the exception of Ni$_2$Ta and
Ni$_4$W where 400 eV and 440 eV respectively were required.  All
structures were tested for mechanical stability, and elastic constants
were not calculated for structures that were found to be mechanically
unstable, though they were included in the enthalpy of formation
plots, as it is possible for the stabilizing distortions to affect the
calculated total energy only weakly.

To quantify bond strength between individual atoms in structures, we
calculate interatomic force constants and the Crystal Overlap Hamilton 
Population (COHP).  To calculate the force constant $k_{\alpha\beta}$ 
between atoms $\alpha, \beta$ at position $\vec{r}_{\alpha,\beta}$, 
separated by a bond in the direction $\hat{\gamma}_{\alpha\beta}$, we 
calculate the Hessian matrix
${\bf H}_{\alpha\beta}=d^{2}E/d\delta\vec{r}_{\alpha}d\delta\vec{r}_{\beta}$.
We use density functional perturbation theory internal to VASP to 
calculate ${\bf H}$ within a supercell of sufficient size that atoms 
lie at least 4.2~\AA~ away from their own repeated images.  Restricting 
our attention to longitudinal (bond stretching) interactions and assuming 
central forces, we define
\begin{equation}
k_{\alpha\beta} \equiv \hat{\gamma}_{\alpha\beta}\cdot{\bf H}_{\alpha\beta}\cdot\hat{\gamma}_{\alpha\beta}
\end{equation}
as the projection of the Hessian along $\hat{\gamma}_{\alpha\beta}$.
$k_{\alpha\beta}$ must be positive for the force to be stabilizing.

The COHP provides an electronic structure-based characterization of
interatomic bond strength~\cite{Dronskowski93}.  One calculates matrix
elements $\langle \alpha L|H|\beta L'\rangle$ of the density
functional theory Hamiltonian between localized atomic orbitals $L$
and $L'$ on a pair of atoms $\alpha$ and $\beta$, then multiplies by
the density of states $N_{\alpha L,\beta L'}$ projected onto the two
orbitals.  We calculate wave functions using a TB-LMTO
method~\cite{Jepsen95} then calculate an {\em integrated} COHP for a
pair of atoms $\alpha\beta$, summing over atomic orbitals and
integrating over energies up to the Fermi energy.

\subsection{Experimental Methods}
To check the validity of our first principles calculations we have
prepared samples of single phase alloys of Ni-Ta by arc melting the
pure element constituents (Ni 99.995\% and Ta 99.95\%) under an argon
atmosphere. The alloys were then suction cast into water cooled copper
molds to form rods of 2mm diameter.  Single phases of the rod
samples were verified by x-ray powder diffraction (XRD)
(Figure~\ref{fig:Ni-Ta_XRD}).  Cylindrical samples were obtained by
sectioning the rods with a diamond saw into 4mm segments. The ends of
the segments were polished to a 3 micron finish.  The elastic
constants of the cylinders were calculated using data obtained from a
Magnaflux Quasar Resonant Ultrasound Spectrometer. RUS involves
placing the cylindrical samples diagonally between two ultrasonic
transceivers and recording the natural modes of vibration~\cite{migliori1997resonant}. 
A Levenberg-Marquardt algorithm is then
used to determine the elastic constants by finding the best fit
solution through minimization of the difference between measured and
calculated natural modes through iterative changes to the values of
elastic constants.

\noindent%
\begin{minipage}{\linewidth}
\makebox[\linewidth]{%
  \includegraphics[trim = 0mm 0mm 0mm 0mm, clip, width=\textwidth]{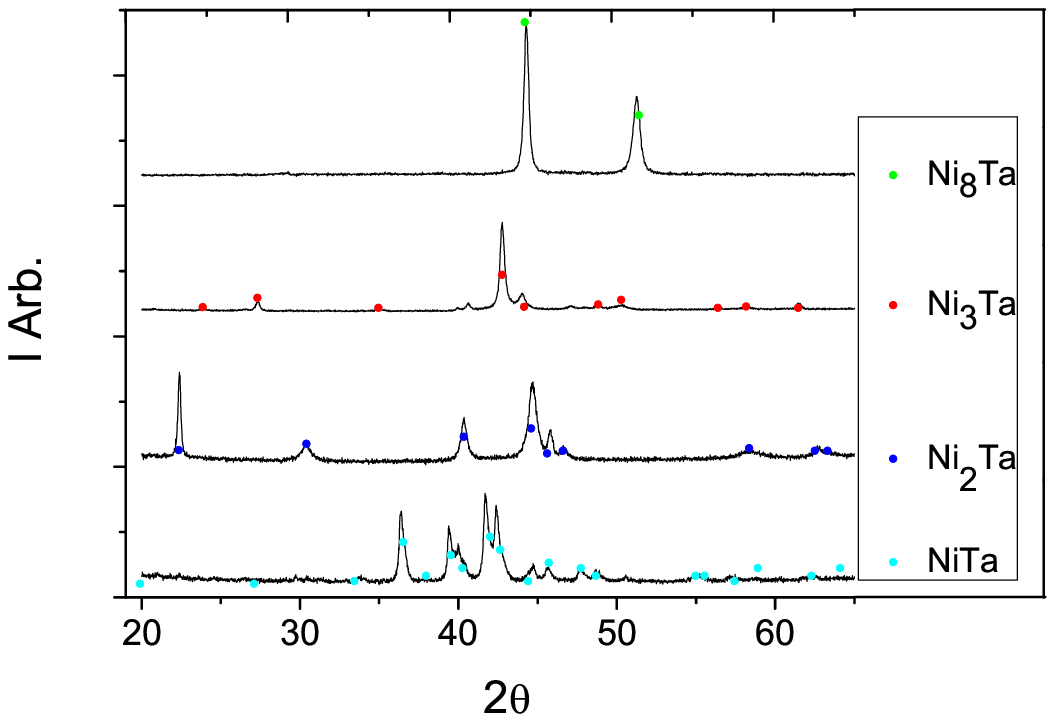}}
\captionof{figure}{XRD verification of single phases of Ni$_{8}$Ta, Ni$_{3}$Ta, Ni$_{2}$Ta,and NiTa.}
\label{fig:Ni-Ta_XRD}
\end{minipage}

The RUS measurement technique is limited to nonmagnetic samples. The 
study using the RUS measurements for comparison to theory is limited to 
the Ni-Ta system where single phases of Ni$_{8}$Ta, Ni$_{3}$Ta, 
Ni$_{2}$Ta and NiTa, which are nonmagnetic, can be 
produced, allowing for RUS measurements. Rods of the pure elements Ni 
and Ta can also be cast, but of the two only Ta produces 
nonmagnetic rods.

\section{Results}

In this section we first discuss atomistic structure, then present
results on thermodynamic stability for each alloy system considered,
finally we address the elastic moduli.

\subsection{Structure}

Because we consider a large number of specific structures, we
establish a numbering scheme to unambiguously identify them (see
Table~\ref{tab:structs}).  In the text we refer to a given
crystalline material using the notation Compound.Pearson
(e.g. Co$_2$Ta.cF24).  Of particular interest are the Frank-Kasper
structures, which we take as amorphous approximants owing to the
prevalence of their coordination polyhedra in many metallic
glasses~\cite{Sheng06,Fang11}.  Canonical Frank-Kasper polyhedra 
have coordination numbers CN=12, 14, 15 and 16.  However, we include 
the Bernal holes~\cite{Bernal1964,Nelson1983}, notably CN=10, as 
tetrahedral but non-canonical Frank-Kasper polyhedra~\cite{Sheng06} 
suitable for smaller atoms.  Also, the tetrahedral close-packing of
the Frank Kasper phases matches the packing properties expected in
bulk metallic glasses~\cite{DBMiracle12}.

\begin{table}
\begin{tabular}{r|rrr|l|l}
Ref. & Prototype & Strukt.  & Pearson& FK?      & CN  \\
1    & Cu        & A1       & cF4    & NA       & 12  \\
2    & W         & A2       & cI2    & NA       & 14  \\
3    & Mg        & A3       & hP2    & NA       & 12  \\
4    & Fe$_6$W$_7$ & D8$_5$ & hR13   & $\mu$    & 12, 14, 15, 16  \\
5    & Al$_2$Cu  & C$_{16}$ & tI12   & non-FK   & 10, 15  \\
6    & AuCu$_3$  & L1$_2$   & cP4    & NA       & 12, 18  \\
7    & BaPb$_3$  &          & hR12   & NA       & 12, 18  \\
8    & MgNi$_2$  & C36      & hP24   & Laves    & 12, 16  \\
9    & MgZn$_2$  & C14      & hP12   & Laves    & 12, 16  \\
10   & MgCu$_2$  & C15      & cF24   & Laves$^*$& 12, 16  \\
11   & Ni$_3$Sn  & D0$_{19}$& hP8    & NA       & 12, 18  \\
12 & Al$_3$Ti  & D0$_{22}$  & tI8    & NA       & 12, 14, 16  \\
13 & Cu$_3$Ti  & D0$_{8}$   & oP8    & NA       & 12, 14, 16  \\
14 & NbPt$_3$  &            & mP16   & NA       & 12, 14, 16  \\
15 & MoSi$_2$  & C11$_b$    & tI6    & NA       & 14  \\
16 & MoNi$_4$  & D1$_a$     & tI10   & NA       & 13, 14  \\
17 & Pt$_8$Ti  &            & tI18   & NA       & 13, 14  \\
18 & MoNi      &            & oP56   & $\delta$ & 12, 14, 15, 16  \\
\end{tabular}
\caption{\label{tab:structs}Structure types considered listing common 
prototype, and Strukturbericht and Pearson notations. Frank-Kasper 
phases list common names. $*$ indicates symmetry-breaking distortion.
``Non-FK'' indicates non-canonical Frank-Kasper phase containing CN=10 Bernal 
Hole.  The final column contains a list of coordination numbers
from Voronoi polyhedra appearing in the structure~\cite{Bernal1964,Nelson1983,Sheng06}.}
\end{table}

\begin{figure*}
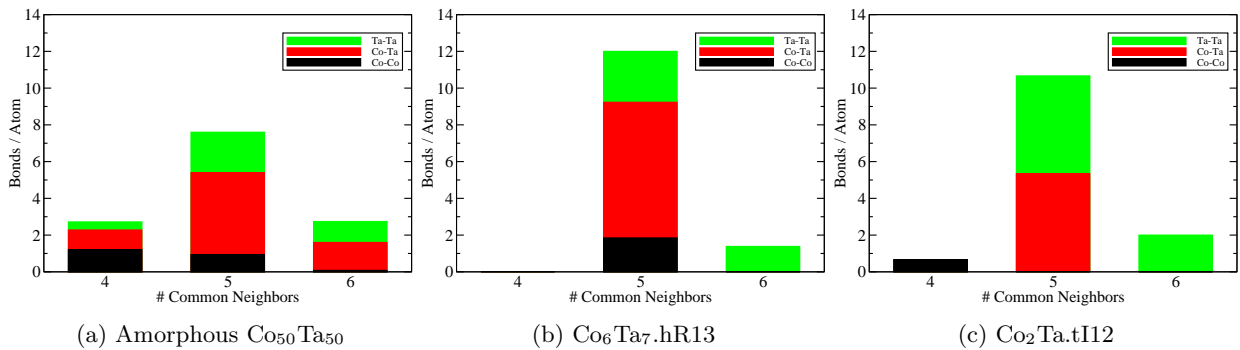

\begin{subfigure}[b]{0.3\textwidth}
\includegraphics[trim = 0mm 0mm 0mm 0mm, clip, width=\textwidth]{Amorphous_HC.eps}
\caption{Amorphous Co\subscript{50}Ta\subscript{50}}\end{subfigure}
\begin{subfigure}[b]{0.3\textwidth}
\includegraphics[trim = 0mm 0mm 0mm 0mm, clip, width=\textwidth]{hR13_HC.eps}
\caption{Co$_6$Ta$_7$.hR13}\end{subfigure}
\begin{subfigure}[b]{0.3\textwidth}
\includegraphics[trim = 0mm 0mm 0mm 0mm, clip, width=\textwidth]{tI12_HC.eps}
\caption{Co\subscript{2}Ta.tI12}\end{subfigure}
\caption{Stacked bar chart of Honeycutt-Andersen common neighbor statistics~\cite{HA87,Ganesh08}.}
\label{fig:HAPairs}
\end{figure*}

To justify use of crystalline phases as amorphous approximants, we
compare Honeycutt-Andersen common neighbor statistics~\cite{HA87,Ganesh08}, 
of crystalline and amorphous structures.  Amorphous structures were 
simulated by quenching 100 atom liquid structures in NPT ensembles from T = 
2500 K down to 300 K over runs of more than 15ps.  All simulations were 
performed at the gamma point with default energy cutoffs.  Shown in Figure~\ref{fig:HAPairs}
is the number of common neighbors between two bonded atoms of given
types. We show results for Co-Ta, but Ni-Ta, Ni-W, and Co-W were also
simulated and generated similar results, with the exception of one
Ni-W amorphous sample which was likely out of equilibrium.

All structures have many bonds with $n=5$ common neighbors, especially
between unlike atomic species, reflecting the prevalence of
icosahedral ordering in Frank-Kasper phases and amorphous materials. Very
few Co-Co bonds have $n=6$ common neighbors, and very few Ta-Ta bonds
have $n=4$ common neighbors, reflecting the relative sizes of Co and
Ta atoms. Because hR13 is a canonical Frank-Kasper phase with no bonds sharing
$n=4$ common neighbors, we utilize tI12 to capture the role of $n=4$
Co-Co bonds.

\subsection{Stability}

\begin{figure*}
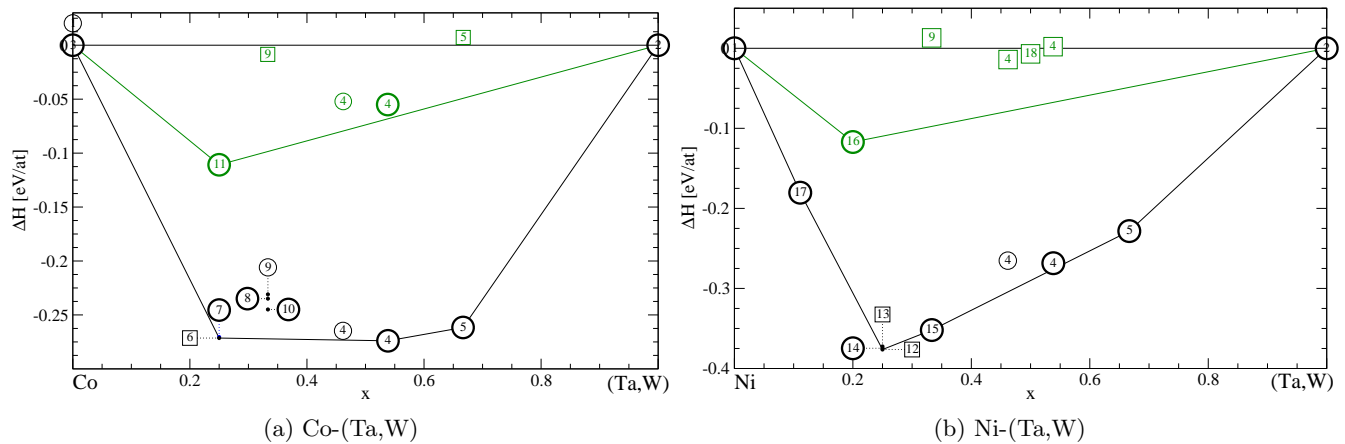

\begin{subfigure}[b]{0.49\textwidth} \includegraphics[trim = 0mm 0mm 0mm 0mm, clip, width=\textwidth]{CoTaW_ConvexHull2.eps}
	\caption{Co-(Ta,W)}
\end{subfigure}
\begin{subfigure}[b]{0.49\textwidth}
	\includegraphics[trim = 0mm 0mm 0mm 0mm, clip, width=\textwidth]{NiTaW_ConvexHull2.eps}
	\caption{Ni-(Ta,W)}
\end{subfigure}
\caption{Enthalpies of formation for (Co,Ni)-(Ta,W) alloys. Values for Ta shown in black, for W shown in green. Plotting symbols explained in text.}
\label{fig:EOF}
\end{figure*}

Figure~\ref{fig:EOF} summarizes our calculated enthalpies of
formation.  Vertices of the convex hull of enthalpy as a function of
composition are predicted to be stable phases at low
temperature~\cite{Mihal04}.  We employ special plotting symbols to
indicate phases claimed experimentally to be stable at low temperature
(heavy circles) and high temperature (light circles). Phases whose
stability or existence is in question are shown as squares.  From the
prevalence of heavy circles on or near the convex hull we see general
(though imperfect) agreement with the experimentally reported phase
diagrams. We briefly summarize our findings for the four alloy systems
of primary interest.

\subsubsection{Co-Ta}
For Co-Ta (Fig.~\ref{fig:EOF}a), at $x=0.25$, we find that cP4 and
hR12 (reference numbers 6 and 7) are nearly degenerate, with cP4
(stability not reported experimentally) favored by 1 meV/atom, which
is closer than DFT can reliably distinguish.  At $x=0.33$, three
different Laves phases have been reported (reference numbers 8-10),
with conflicting claims of stability and uncertain composition.  We
find that none of these phases lies on the convex hull. Further, all
of their structures are mechanically unstable to deformation, casting
doubt on the reported structure and stability of this phase.  In our
plot, we show the energy of a distorted cF24 structure which is
mechanically stable, for comparison with the undistorted hP12 and
hP24.  In the vicinity of $x=0.5$ lies the Frank-Kasper $\mu$ phase.
This phase is common to many alloy systems containing fourth and fifth
row early transition metals with third row late transition metals.  In
most cases the phase shows a broad composition range at high
temperature but favors an ETM-rich low temperature limit (i.e. with
the mixed occupancy $3a$ site occupied by an ETM).  This feature is
correctly reproduced by our calculation.  At $x=0.67$, the tI12 phase
(reference number 5) is a non-canonical Frank-Kasper phase, as it
contains a CN=10 Bernal Hole, in addition to a canonical CN=15 Kasper
polyhedron.

\subsubsection{Ni-Ta}
For Ni-Ta (Figure \ref{fig:EOF}b) at $x=0.11$ we find
Ni\subscript{8}Ta.tI18 to be low-temperature stable.  The experimental
phase diagram has a tI36 structure stable, however no crystallographic
refinement exists, so we use the known tI18 phase instead.  There is
disagreement on recent phase diagrams concerning the stable structure
at $x=0.25$, and we find three different structures have nearly
identical enthalpies (tI8 is the lowest).  The main source of disagreement
between our T = 0 K phase diagram and experimental phase diagrams is the
stability of NiTa.hR13, with even the ETM-rich variant lying 7
meV/atom above the convex hull.  Fig.~\ref{fig:Ni-Ta_XRD} shows the
diffraction patterns of our experimentally cast rods, verifying the
existence of the expected phases in our own samples.

\subsubsection{(Co,Ni)-W}
Our calculations verify the known Co$_3$W and Ni$_3$W phase stabilities.  
However, the reported Co$_7$W$_6$ phase lies above the convex hull, and 
additionally favors the ETM-rich limit contrary to experimental report.  
This phase has not been reported in the Ni-W alloy system, and we indeed 
find it lies well above the convex hull.  However, we will study the 
electronic and elastic properties of this hypothetical phase in order 
to elucidate trends with respect to composition.  Notice that enthalpies 
of formation for alloys with W are lower than enthalpies with Ta.  This 
does not necessarily reflect lower mechanical stability or melting points 
for the compounds, as the greater cohesive energy of tungsten compared 
to tantalum contributes to a reduction of the alloy formation enthalpies.  
An equiatomic Ni-W phase with orthorhombic symmetry has been observed 
low temperature stable~\cite{JMWalsh73}, however no atomic structural data exists.  Owing 
to similar chemical identity and Bravais lattice, we attempted to use the Frank-Kasper phase
MoNi.oP56 with Mo substituted for W, but found that this structure lies 
66 meV/atom above the convex hull and likely is not the correct phase.  
It was also too computationally expensive to compute elastic moduli
for this phase.

\subsection{Binary Elastic Moduli}
We examine the effect of alloying on the elastic moduli by using CPA
to approximate a hypothetical alloy where no interspecies interactions
exist.  That is, for a well-ordered phase with stoichiometry
A\subscript{x}B\subscript{1-x}, we compare its elastic moduli to a
hypothetical solution of pure specie A and pure specie B with a
stoichiometric ratio x:(1-x) in the CPA approximation.  As input for
CPA, we use our computed elemental elastic moduli.  These agree
closely with experimental moduli for Ta and W but are relatively high
for Co and Ni.  Note that our calculation are valid at 0 K, while the
experimental values are room temperature, so it is expected that our
values should be high, especially for non-refractory elements.  As the
CPA approximation only incorporates elemental elastic moduli with no
atomic environmental details, the deviation of computed
polycrystalline moduli from the CPA approximation yields a measure of
the relative importance of atomic environment and alloying species for
elastic moduli.

\begin{figure*}
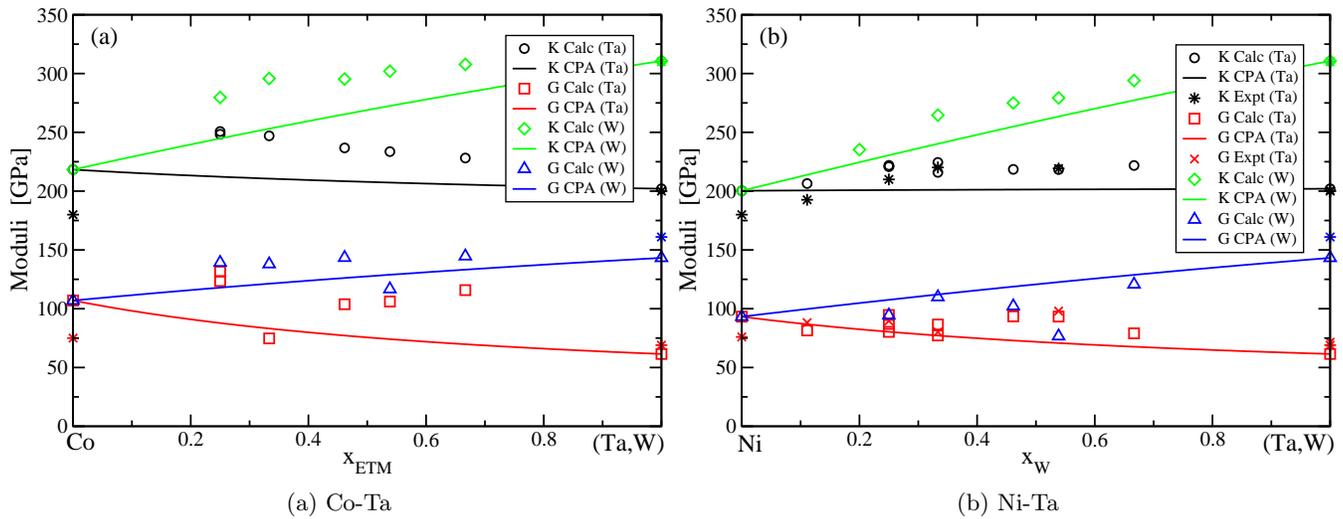

\begin{subfigure}[b]{0.49\textwidth}
	\includegraphics[trim = 0mm 0mm 0mm 0mm, clip, width=\textwidth]{CoETM_moduli.eps}
	\caption{Co-Ta}
\end{subfigure}
\begin{subfigure}[b]{0.49\textwidth}
	\includegraphics[trim = 0mm 0mm 0mm 0mm, clip, width=\textwidth]{NiETM_moduli.eps}
	\caption{Ni-Ta}
\end{subfigure}
\caption{Elastic moduli for (Co,Ni)-(Ta,W) compounds. Lines show CPA 
estimates. Data points are our DFT calculations. Asterisks are 
experimental values from our experiments on Ni-Ta alloys and from the 
literature for pure elements.}
\label{fig:KG}
\end{figure*}



All crystal structures are elastically anisotropic, and it is of interest to characterize 
the anisotropy of our amorphous approximants.  We define three anisotropies~\cite{Nye85,Tromans11}: 
$A_Z$ (Zener), $A_G$ (shear) and $A_E$ (Young's) as
$$
A_Z = \frac{2C_{44}}{C_{11}-C_{12}},
$$
$$
A_G=\frac{S_{44}+S_{66}}{2S_{44}},
$$
and
$$
A_E=\frac{S_{11}}{S_{33}},
$$
all of which are 1 for isotropic structures, where $S_{ij}$ are elements of the compliance matrix. 
Table~\ref{tab:aniso} compares calculated anisotropies of our hR13 and tI12 structures with the
four pure elements.  Our anisotropies are close to one, similar to those seen in the pure metals, 
with hR13 exhibiting less anisotropy than tI12.  These anisotropies can be taken as estimates of 
the local anisotropy expected at the atomic level in actual amorphous structures.  Recall that 
the polycrystalline averages are expected to reflect the globally isotropic properties of the bulk 
amorphous structures.

\begin{table}
\begin{tabular}{c|c|c|c|c}
Composition & Structure & $A_Z$	 & $A_G$ & $A_E$\\
\hline
Co-Ta & hR13 & 0.70 & 0.85 & 1.13 \\
Ni-Ta & hR13 & 0.81 & 0.91 & 1.14 \\
Co-W  & hR13 & 0.76 & 0.88 & 1.18 \\
Ni-W  & hR13 & 1.12 & 1.06 & 1.45 \\
\hline
Co-Ta & tI12 & 0.90 & 1.17 & 1.27 \\
Ni-Ta & tI12 & 1.32 & 1.24 & 1.50 \\
Co-W  & tI12 & 1.24 & 1.12 & 1.20 \\
Ni-W  & tI12 & 1.42 & 1.16 & 1.07 \\
\hline
Co    &  hP2 & 1.01 & 1.00 & 1.31 \\
Ni    &  cF4 & 2.18 & 1 & 1 \\
Ta    &  cI2 & 1.16 & 1 & 1 \\
W     &  cI2 & 0.83 & 1 & 1 \\
\end{tabular}
\caption{Measures of anisotropy for materials of interest.  $A_{G} = A_{E} = 1$ for cubic structures.}
\label{tab:aniso}
\end{table}

Shown in Figure \ref{fig:KG} are our calculated K values, compared
with CPA estimates.  All CPA estimates are indicated by lines, all
calculated moduli by individual data points, and for Ni-Ta asterisks
indicate experimental results.  Our calculated Ni-Ta bulk moduli show
excellent agreement with our experimental results.  For all four
chemical families, CPA gives reasonable estimates for bulk moduli,
with at most a 16\% deviation between estimated and calculated bulk
moduli.  However, in all alloy systems and for all structures
examined, CPA underestimates the bulk modulus.  This suggests the
dominant correction to the bulk moduli is chemical bonding and not
atomic environmental details such as the prevalence of tetrahedra.
Shear moduli show relatively larger and less regular deviations from
CPA, suggesting that bond topology plays a significant role.
Nonetheless, our calculated Ni-Ta shear moduli are in good agreement
with experiment (crosses).
 
\begin{table}
        \begin{tabular}{|c|c|c|c|c|}
        \hline
        & $\Delta$v & $\Delta$h & $\Delta$K & $\Delta$G  \\
        \hline
        $\Delta$v &       &  0.20 & -0.83 & -0.66 \\
        \hline
        $\Delta$h &  0.20 &       &  0.06 & -0.47 \\
        \hline
        $\Delta$K & -0.83 &  0.06 &       &  0.60 \\
        \hline
        $\Delta$G & -0.66 & -0.47 &  0.60 &       \\
        \hline
        \end{tabular}
        \caption{Correlation coefficients for linear 
regressions involving elastic moduli and thermodynamic quantities of 
interest, where rows denote independent variables and columns dependent 
variables for a given regression.}
        \label{tab:correlations}
\end{table}

Shown in Table \ref{tab:correlations} are the correlation coefficient
for various linear regressions across all calculated CoTa, CoW, NiTa,
and NiW alloys, where the sign of the correlation coefficient denotes
the sign of the slope.  Here $\Delta$K and $\Delta$G (units GPa) are
deviations of calculated bulk and shear moduli from CPA estimates,
$\Delta$v (units \AA$^{3}$ per atom) the deviation of volume per atom
from a linear interpolation of pure elements, and $\Delta$h (units eV
per atom) the enthalpy of formation per atom as illustrated in
Fig.~\ref{fig:EOF}.  Correlations of elastic moduli with $\Delta$v and
$\Delta$h reflect structure and bonding effects that are missing from
CPA.  There is only a weak correlation between $\Delta$h and $\Delta$v, 
though the associated slope is positive, expected as 
increased bond strength (more negative $\Delta$h) draws
atoms closer together (more negative $\Delta$v).  $\Delta$G and
$\Delta$K are both correlated to $\Delta$v, with $\Delta$K in
particular strongly correlated.  This is line with other work that
shows that K and G correlate with V and that the correlation for K is
particularly strong~\cite{Wang12}.  This is also true for individual
chemical families, and for $\Delta$K all chemical families' 
regressions have similar slopes.  This strong correlation of
$\Delta$K with $\Delta$v explains why all structures have CPA
underestimating the bulk moduli (positive $\Delta$K), as all
structures were also observed to have negative $\Delta$v, as expected.
The slopes of $\Delta$G and $\Delta$K are both negative, as decreasing
$\Delta$v draws atoms closer together, shortening bonds and enhancing
interatomic force constants.  The observed correlation of $\Delta$G
with $\Delta$K is likely due to the underlying correlation of each
with $\Delta$v. 
 
A goal of metallic glass design is to predict glass-forming compounds
with high ductility.  Thus, as a guide, we plot the Poisson ratio's of
the various alloys under discussion.  Our computed T = 0 K crystalline
Poisson ratios are expected to be systematically low relative to the
corresponding glasses, as G decreases more rapidly than K as
temperature increases~\cite{XFLiu10}, and amorphous G and K are lower relative to
crystalline values by around 30\% and 10\%, respectively.  Here we see
no systematic trend in the choice of Ta versus W (empty versus filled
plotting symbols) but Ni generally has higher Poisson's ratio than Co
(red versus blue).  Empirically, it has been observed that $\nu =
0.32$ serves as a rough criterion for separating ductile and brittle
behavior in amorphous materials~\cite{JJLewandowski05}.  The majority
of our Ni alloys lie above this criterion and Co alloy below.  In
particular, all Ni-W alloys satisfy this criterion, and Ni-W in the
amorphous approximant structure hR13 shows particularly large
Poisson's ratios.  Combined with the large bulk modulus due to the
presence of tungsten, we propose that Ni-W is a candidate system for
future research into strong amorphous materials with high ductility.
 
\begin{figure}
\includegraphics[trim = 0mm 0mm 0mm 0mm, clip, width=3.5in]{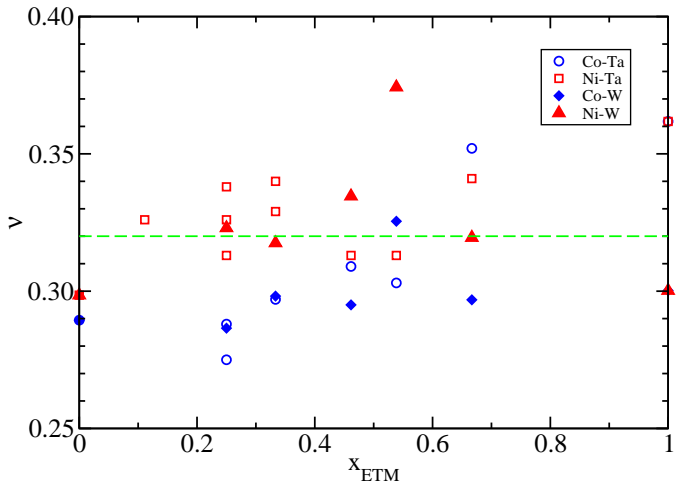}
\caption{\label{fig:nu}Calculated Poisson ratios. Dashed line at $\nu$=0.32
is putative threshold for ductility.}
\end{figure}

\section{Analysis and Discussion}
To understand trends in elastic constants of these alloys we now look
into the interatomic interactions.  Within a first principles approach
there is no unique decomposition of interactions into pairwise and
many-body forces, and no simple notion of a chemical bond, especially
for metals.  However some heuristic measures are available.  Here we
examine the interatomic force constants, which can be regarded as
springs connecting the atoms, and the crystal overlap Hamilton
populations (COHPs) which are a measure of the covalency of electronic
wave functions.

\subsection{Force Constants and COHP}
\begin{table*}
\begin{tabular}{r|lll|lll|lll|rrr}
Compound & 
\multicolumn{3}{c|}{ETM-ETM (44x)}&
\multicolumn{3}{c|}{ETM-LTM (32x)}&
\multicolumn{3}{c|}{LTM-LTM (4x)} & mean & mean & \\
tI12     & R   & k  & $\rho_{\rm COHP}$  & R   & k  & $\rho_{\rm COHP}$  & R   & k  & $\rho_{\rm COHP}$ & k   & $\rho_{\rm COHP}$ & K\\
\hline 
NiTa$_2$ & 3.12& 4.2& 0.31  & 2.63& 4.7& 0.25  & 2.43& 4.2& 0.025 &30.7 & 0.27 & 222 \\
CoTa$_2$ & 3.11& 5.2& 0.31  & 2.61& 3.7& 0.30  & 2.48& 4.0& 0.030 &32.0 & 0.29 & 228 \\
NiW$_2$  & 3.02& 4.0& 0.31  & 2.55& 7.6& 0.34  & 2.38& 4.3& 0.041 &37.7 & 0.31 & 294 \\
CoW$_2$  & 3.02& 5.1& 0.37  & 2.54& 6.7& 0.36  & 2.38& 4.8& 0.037 &39.9 & 0.35 & 308 \\
\end{tabular}
\caption{\label{tab:RkC}Data for tI12. R are average bond lengths for 
nearest neighbors under 4~\AA. $k$ are averaged over atoms 
(i.e. weighted by the number of bonds per atom) in units eV/\AA$^2$, 
$\rho_{\rm COHP}$ is iCOHP volume density in units eV/\AA$^3$, and $K$ is the 
calculated bulk moduli in units GPa.}
\end{table*}

To compare different ETM and LTM substitutions, Table \ref{tab:RkC}
shows k, the mean force constants for the near neighbor bonds of a
given species combination, $\rho_{\rm COHP}$, the total iCOHP per unit
volume for bonds of a given species combination up to 4 \AA, and K, 
the bulk moduli in the structural prototype tI12.  To calculate the 
mean force constant, we sum force constants for all bonds up to 4 \AA~ 
for a given cell then divide by the number of atoms.

Both the mean force constant and $\rho_{\rm COHP}$ correlate with the
bulk moduli.  This is especially notable in the mean force constant,
where there is a large increase in mean force constant performing a Ta
$\rightarrow$ W substitution and a relatively small increase
performing a Ni $\rightarrow$ Co substitution, but the effect is also
present in $\rho_{\rm COHP}$.  As a force constant gives a measure of
the stiffness of an individual bond, this mean force constant gives a
measure of the total stiffness of all bonds, and bulk modulus is
increased under chemical substitution by an overall increase in the
interatomic force constant.  We also see in Table \ref{tab:RkC} that
performing a Ni $\rightarrow$ Co or a Ta $\rightarrow$ W substitution
enhances $\rho_{\rm COHP}$.  Thus these substitutions have enhanced 
the bonding nature of the electronic states.

\begin{figure*}
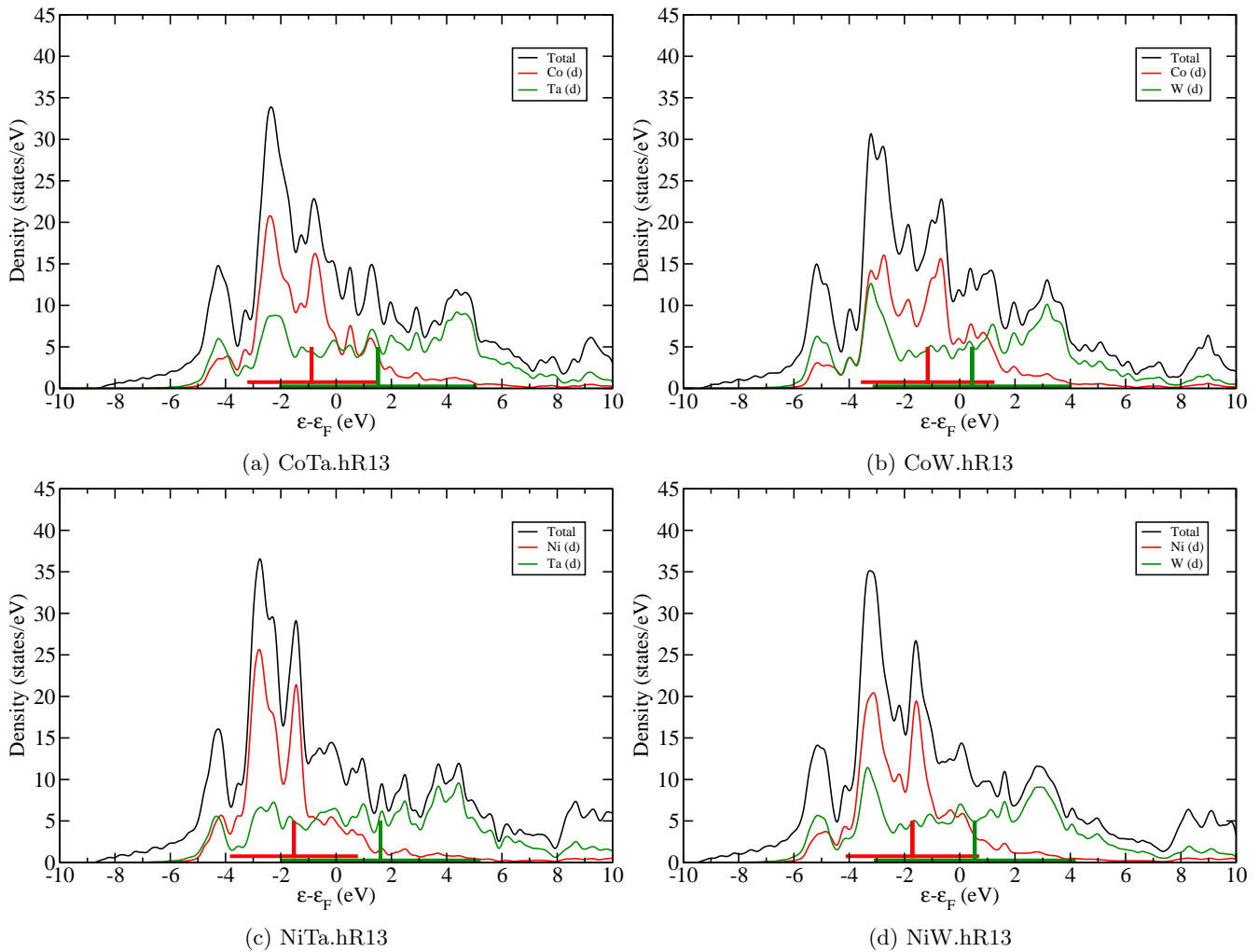

\begin{subfigure}[b]{0.49\textwidth}
	\includegraphics[trim = 0mm 0mm 0mm 18mm, clip, width=\textwidth]{CoTahR13+3TaDOS-width.eps}
	\caption{CoTa.hR13}
\end{subfigure}
\begin{subfigure}[b]{0.49\textwidth}
	\includegraphics[trim = 0mm 0mm 0mm 18mm, clip, width=\textwidth]{CoWhR13+3WDOS-width.eps}
	\caption{CoW.hR13}
\end{subfigure}
\begin{subfigure}[b]{0.49\textwidth}
	\includegraphics[trim = 0mm 0mm 0mm 18mm, clip, width=\textwidth]{NiTahR13+3TaDOS-width.eps}
	\caption{NiTa.hR13}
\end{subfigure}
\begin{subfigure}[b]{0.49\textwidth}
	\includegraphics[trim = 0mm 0mm 0mm 18mm, clip, width=\textwidth]{NiWhR13+3WDOS-width.eps}
	\caption{NiW.hR13}
\end{subfigure}
\caption{Electronic densities of states for hR13 structures. Line segments indicate the mean and standard deviations of the element projected $d$-bands.}
\label{fig:hR13eDOS}
\end{figure*}

To further understand the enhancement of bonding, we calculate
electronic densities of state (Fig.~\ref{fig:hR13eDOS}).  The
low-energy peak near -4 or -5 eV consists of $sd$-hybrid orbitals,
followed by a series of higher-energy peaks consisting solely of $d$
orbitals, with the Fermi level lying in the middle of the ETM $d$-band
and at or above the top of the LTM $d$-band.  For Co-W, the Co and W
$d$ bands are closely aligned, inducing strong hybridization of Co and
W $d$ orbitals.  This effect is present in all structures we have
examined.  Performing a W $\rightarrow$ Ta substitution shifts the ETM
$d$-band up relative to the LTM $d$-band, decreasing the $d$-band
overlap and diminishing hybridization.  Performing a Co $\rightarrow$
Ni substitution shifts the LTM $d$-band down relative to the ETM
$d$-band, also decreasing the $d$-band overlap.  Both of these induce
an decrease in the hybridization of the ETM-LTM $d$-bands.  As
hybridization generally creates bonding states below the Fermi level,
this reduction in hybridization going from W $\rightarrow$ Ta and Co
$\rightarrow$ Ni decreases the overall bonding characteristic of the
occupied states, leading to the observed trends in $\rho_{\rm COHP}$,
and hence in bulk modulus.

\subsection{Microstructural Details:  Ternaries and Quaternaries}

Elemental properties provide the dominant contribution to the
elasticity of these ETM-LTM intermetallic compounds, as can be seen in
the qualitative agreement of calculated alloy moduli with the CPA
predictions shown in Fig.~\ref{fig:KG}.  In Table~\ref{tab:Averages}
we see that the small decreases in modulus from Co to Ni, and the
large increases from Ta to W, are echoed in the moduli of the hR13
Frank-Kasper structure.

\begin{table}
\begin{tabular}{l|lrr|lrr}
&            & K   & G  &              & K   & G \\
\hline
&         Co & 218 &107 &          Ni  & 200 & 93 \\
\hline
Nb (K=173, G=22) &
Co$_6$Nb$_7$ & 213 & 92 & Ni$_6$Nb$_7$ & 198 & 81 \\
Ta (K=202, G=61) &
Co$_6$Ta$_7$ & 234 &106 & Ni$_6$Ta$_7$ & 218 & 93 \\
W (K=331, G=143) &
Co$_6$W$_7$  & 302 &117 & Ni$_6$W$_7$  & 279 & 77 \\
\end{tabular}
\caption{\label{tab:Averages}Elastic moduli for pure elements and binary hR13 structures.}
\end{table}

While binary amorphous metals exist, size mismatch criteria and
material property tuning favor the usage of multiple constituent
species in amorphous metals for practical applications, and thus the
question of transferability of binary results to structures with 3 or
more constituent species must be addressed.  In addition, there is
still the lingering need to quantify how atomic environment affects
the elastic moduli.  To answer both these questions, we perform
chemical substitutions in a binary structure to yields ternaries and
quaternary structures.  This changes the chemical identities of
formerly equivalent sites, altering local chemical ordering.
 
Shown in Table \ref{tab:QuathR13} is a comparison of binary hR13
structures (including also alloys with Nb, an ETM) with nearly equiatomic
composition quaternary variants of hR13 and six associated
ternaries. Site occupancies in the quaternary has been chosen to
maintain the ETM/LTM nature of sites and minimize energy, and the
ternaries were formed by keeping the early/late site identity fixed.

To compare our binary results to ternaries and quaternaries, we here
use a simple chemical environment averaging scheme between ETM and
LTM, with a equiatomic ABCD mixture with A and B LTM and C and D ETM
approximated by 1/4*(AC+AD+BC+BD), and an ABC mixture (with C having
near 50\% concentration) approximated by 1/2*(AC+BC).  Here AC, AD, BC
and BD refer to the relevant binary hR13 structure with the associated
chemical formula.  As an example, the predicted bulk moduli of
Co\subscript{6}Ta\subscript{3}W\subscript{4} would be the average bulk
modulus of Co\subscript{6}Ta\subscript{7} and
Co\subscript{6}W\subscript{7}.  While this ignores interspecies
ETM-ETM and LTM-LTM bonds present (i.e. AB and CD), binary enthalpies
of formation for ETM-ETM and LTM-LTM families are weak compared to
ETM-LTM families, suggesting that as a first approximation we may
assume the differences in interspecies ETM-ETM and LTM-LTM bond
strength average out.

Differences between our predicted interpolated elastic moduli and
computed elastic moduli follow the trends previously reported for CPA.
Again we see bulk moduli negligibly affected by atomic environment and
predominately determined by the alloying species, with deviations in
bulk moduli below 2.6\% for all structures.  For shear moduli, the
structures can be placed into two categories: those structures that
have only one ETM species or else two ETM species from the same group
(here Nb and Ta belong to group IV) which have deviations in shear
moduli below 3.7\%, and those that have ETM species from different
groups (here W from group V together with Nb or Ta from group IV)
which have deviations in shear moduli between 10.0\% and 18.7\%.  In
all cases where predicted shear moduli deviate significantly from
calculated shear moduli, the computed shear moduli have been enhanced.

\begin{table}
\begin{tabular}{ |c|c|c|c| }
\hline
Chemical Formula & Calculated K, G & Averaged K, G & Deviation \\
\hline
Co\subscript{3}Ni\subscript{3}Nb\subscript{7} & 205,  87 & 205,  87 & 0.3, 0.5 \\
\hline
Co\subscript{3}Nb\subscript{3}Ta\subscript{4} & 225, 100 & 223,  99 & 0.7, 1.4 \\
\hline
Co\subscript{3}Ni\subscript{3}Ta\subscript{7} & 225, 102 & 226, 100 & 0.4, 2.2 \\
\hline
Ni\subscript{6}Nb\subscript{3}Ta\subscript{4} & 210,  89 & 208,  87 & 0.9, 2.2 \\
\hline
Co\subscript{3}Ni\subscript{3}W\subscript{7}  & 290,  94 & 291,  97 & 0.3, 2.4 \\
\hline
Co$_3$Ni$_3$Nb$_3$Ta$_4$ & 217, 97 & 216, 93 & 0.1, 3.7 \\
\hline
\hline
Co\subscript{6}Ta\subscript{4}W\subscript{3} & 261, 123 & 268, 111 & 2.5, 10.0 \\
\hline
Co\subscript{3}Ni\subscript{3}Ta\subscript{3}W\subscript{4} &  253, 115 & 258, 98 & 2.1, 14.6 \\
\hline
Co$_6$Nb$_3$W$_4$ & 261, 121 & 257, 104 & 1.5, 16.4 \\
\hline
Ni$_6$Nb$_3$W$_4$ & 241,  92 & 239, 79 & 0.8, 16.8 \\
\hline
Ni$_6$Ta$_4$W$_3$ & 242, 100 & 249, 85 & 2.6, 17.8 \\
\hline
Co$_3$Nb$_3$Ni$_3$W$_4$ & 248, 109 & 248, 92 & 0.2, 18.7 \\
\hline
\end{tabular}
\caption{Comparison of calculated ternary and quaternary hR13 moduli with averaged values of binaries. Moduli are given in GPa, and deviations in percentages.}
\label{tab:QuathR13}
\end{table}
 
That mixing Co and Ni or Nb and Ta causes little deviation in shear
modulus, but mixing Ta and W does, is further evidence for the
dependence of shear modulus on atomic environment.  Co and Ni have
similar atomic radii and electronegativity, as do Nb and Ta.  For a
topologically close packed structure like hR13, substitution of these
chemical species should not noticeably affect bond lengths and ionic
charges, yielding similar calculated and averaged results.  However,
Nb and Ta have larger atomic radii and lower electronegativity than W,
leading to larger charge transfers and changes in bond length,
reducing the accuracy of our averaging scheme while generally
increasing bonding strength.

\section{Conclusions}

In this paper we examine the elasticity of various early transition
metal-late transition metal crystalline binary alloys using first
principles calculations and comparison with various averaging schemes.
Calculated bulk moduli were reasonably close to the coherent potential
approximation using pure elemental species, but CPA predictions were
systematically low.  This deviation correlates strongly with volume
per atom.  Larger and less regular deviations were observed for shear
moduli, suggesting structural distortion being responsible for the
deviation.  Select ternary and quaternary structures were examined and confirmed
these trends.  To explain the dependence of elastic moduli on chemical 
bonding, force constants and electronic densities of state were 
calculated and it was found early transition metals are responsible 
for the strongest bonding, which agrees with observed trends in the 
bulk moduli.  We find that Ni-W alloys have the largest Poisson ratios 
among the compositions studied and hence hold promise as the basis for 
design of ductile metallic glasses.

\section{Acknowledgements}
We would like to give special thanks to DOD-DTRA for funding this research 
under contract number DTRA-11-1-0064.

\end{document}